\documentclass[a4paper,11pt, headings=big,DIV=12]{scrartcl}
\usepackage{amsmath,amssymb}
\usepackage{color,xcolor}
\definecolor{blue}{rgb}{0,0,0.5} 
\usepackage[colorlinks,linkcolor=blue,urlcolor=blue,citecolor=blue]{hyperref}
\usepackage{graphicx}
\addtokomafont{disposition}{\rmfamily\boldmath}
\usepackage[utf8]{inputenc}
\usepackage{slashed}
\usepackage{multirow}
 \usepackage{bbold}
 \usepackage{mathbbol}

\pdfoutput=1 

\usepackage{bbm}

\usepackage[utf8]{inputenc}
\allowdisplaybreaks
\DeclareMathAlphabet{\mathantt}{OT1}{antt}{li}{it}
\DeclareMathAlphabet{\mathpzc}{OT1}{pzc}{m}{it} 

\usepackage{simplewick} 

\usepackage{makecell}	
\usepackage{tabu}		

\usepackage[makeroom]{cancel} 

\usepackage[normalem]{ulem} 

\usepackage{amsthm}

\usepackage[utf8]{inputenc}

\usepackage{amsmath,amsfonts,amsthm,amssymb} 

\numberwithin{equation}{section}

\newcommand{\vev}[1]{\langle #1 \rangle}

\newcommand{\p}{\partial}
\newcommand{\para}{\parallel}

\newcommand{\eq}{&\quad}

\newcommand{\rig}{\right.}
\newcommand{\lef}{\left.}

\newcommand{\hp}{{\hat{\phi}}}
\newcommand{\hc}{{\hat{\chi}}}

\newcommand{\diag}{\text{diag}}
\newcommand{\eff}{\text{eff}}

\newcommand{\tr}{\text{tr}}

\newcommand{\mcm}{\mathcal{M}}
\newcommand{\mco}{\mathcal{O}}

\newcommand{\mbi}{\mathbb{1}}
\newcommand{\mbr}{\mathbb{R}}

\newcommand{\al}{\alpha}
\newcommand{\be}{\beta}
\newcommand{\ch}{\chi}
\newcommand{\de}{\delta}
\newcommand{\e}{\epsilon}
\newcommand{\ph}{\phi}
\newcommand{\g}{\gamma}

\newcommand{\la}{\lambda}
\newcommand{\m}{\mu}

\newcommand{\s}{\sigma}

\newcommand{\x}{\xi}

\newcommand{\D}{\Delta}
\newcommand{\Ph}{\Phi}
\newcommand{\G}{\Gamma}
\newcommand{\La}{\Lambda}

\newcommand{\X}{\Xi}

\begin{document}

 \thispagestyle{empty}

\begin{flushright}
\begin{tabular}{l}
 UUITP- 50/20
\end{tabular}
\end{flushright}
\vskip1.5cm

\begin{center}
{\Large\bfseries \boldmath Spontaneous symmetry breaking in free theories with boundary potentials}\\[0.8 cm]
{\Large%
Vladimír Procházka, Alexander Söderberg
\\[0.5 cm]
\small
 Department of Physics and Astronomy, Uppsala University,\\
Box 516, SE-75120, Uppsala, Sweden 
} \\[0.5 cm]
\small
E-Mail:
\texttt{\href{mailto:vladimir.prochazka@physics.uu.se}{vladimir.prochazka@physics.uu.se}},
\texttt{\href{mailto:alexander.soderberg@physics.uu.se}{alexander.soderberg@physics.uu.se}}.
\end{center}

\bigskip

\pagestyle{empty}
\begin{abstract}
Patterns of symmetry breaking induced by potentials at the boundary of free $O(N)$ models in $d=3- \epsilon$ dimensions are studied. We show that the spontaneous symmetry breaking in these theories leads to a boundary RG flow ending with $N - 1$ Neumann modes in the IR. The possibility of fluctuation-induced symmetry breaking is examined and we derive a general formula for computing one-loop effective potentials at the boundary. Using the $\epsilon-$expansion we test these ideas in an $O(N)\oplus O(N)$ model with boundary interactions. We determine the RG flow diagram of this theory and find that it has an IR-stable critical point satisfying conformal boundary conditions. The leading correction to the effective potential is computed and we argue the existence of a phase boundary separating the region flowing to the symmetric fixed point from the region flowing to a symmetry-broken phase with a combination of Neumann and Dirchlet boundary conditions.
 \\

\end{abstract}

\newpage 

\setcounter{tocdepth}{3}
\setcounter{page}{1}
\pagestyle{plain}

\newtheorem{defin}{Definition}
\newtheorem{thm}{Theorem}
\newtheorem{cor}{Corollary}
\newtheorem{pf}{Proof}
\newtheorem{nt}{Note}
\newtheorem{ex}{Example}
\newtheorem{ans}{Ansatz}
\newtheorem{que}{Question}
\newtheorem{ax}{Axiom}

\section{Introduction}

Spontaneous breaking of global symmetries is one of the most universally used tools to understand phase transitions in modern theoretical physics. In this paper we would like to consider its application to systems described by scalar field theories existing on a manifold with a boundary.  A lot has already been understood in the condensed matter context \cite{Diehl:1996kd}, where such systems describe polymer absorption by walls \cite{DE87}. Other than the usual order-disorder phase transition in the bulk (called the ordinary transition), there is a possibility of an extraordinary phase transition at the boundary above the bulk critical temperature. Field theoretically such systems are represented by an $O(N)$ model in $d = 3$ dimensions with polynomial interactions in the bulk where the extraordinary phase transition is triggered by a negative 'boundary mass' term. This representation makes them amenable to study with the techniques of high-energy physics. In particular the machinery of boundary conformal bootstrap \cite{1210.4258} allows for high precision evaluation of correlation functions at the Wilson-Fisher (WF) fixed point (FP) \cite{1808.08155}, which was recently used in evaluation of layer susceptibility at the extraordinary transition point \cite{Shpot:2019iwk, Dey:2020lwp}. Alternatively a wealth of information on these models can be obtained by coupling them to a curved background and calculating the resulting partition function \cite{Herzog:2020lel, Giombi:2020rmc}.

\quad In this work we would like to examine the case when bulk couplings are turned off and instead we include interactions at the boundary. For $d=3- \epsilon$ this still leads to a non-trivial RG flow at the boundary with an interacting infrared (IR) FP, which was recently studied in \cite{Prochazka:2019fah} and \cite{Giombi:2019enr}. In condensed matter literature, scalar models with boundary interactions were considered  long before in \cite{ED88}. In the context of polymer physics, tuning the bulk couplings to zero means considering a rather non-realistic example with two-body monomer interactions confined to the boundary. 

\quad In the realm of high energy physics there are nevertheless important examples of free theories with boundary interactions. For $d=2$ free bosons with boundary potentials have been studied in the context of open strings \cite{Callan:1994ub, Harvey:2000na}. More recently there has been a progress in constraining free scalar theories with boundaries and defects with $d > 2$ by using conformal boostrap techniques \cite{Lauria:2020emq, Behan:2020nsf}.

\quad Finally let us note that free models are often related to interacting ones via dualities such as bosonisation in $d = 2$ dimensions or more refined dualities that have recently been discovered in $d = 3$ dimensions \cite{Karch:2016sxi, Seiberg:2016gmd}. Thus it is possible that already by studying the models that are free in the bulk we can learn something about the interacting theories and their boundary deformations via the duality.

\quad In this paper we would like to consider giving a vacuum expectation value (v.e.v.) to a boundary field. This is not a new idea in itself. In the condensed matter context (cf. \cite{Diehl:1996kd}) this phenomenon gives rise to new kinds of phase transitions called special and extraordinary. These transitions cannot be deduced from the knowledge of the bulk phase diagram itself and are described by a set of independent boundary parameters (couplings, v.e.v.'s, etc.). When the bulk is free there are no bulk parameters to tune so all the non-trivial dynamics happens at the boundary either through edge degrees of freedom or dynamical boundary conditions (b.c.'s). We would like to study the latter in the present work and convince the reader that such a simple set-up can lead to rich physics similar to the phase structure of the Ising model.

\quad Let us start by introducing the class of models we want to work with. We will consider a free $O(N)$ with a boundary potential 
\begin{equation} \label{our model}
\begin{aligned}
S[\ph] = \int_{\mathbb{R}^d_+}d^dx \left[ \frac{(\p_\m\ph)^2}{2} + \de(x_\perp) V(\ph) \right] \ , \quad d = 3 - \e \ ,
\end{aligned}
\end{equation}

where we have suppressed the index notation for $\phi \equiv  \ph^i$ with $i$ running from $1$ to $N$ and used the Euclidean space conventions. The bulk theory has an $O(N)$ symmetry 
\begin{equation}
\phi \to R \phi  \ , \quad R \in O(N) \ ,
\end{equation}

and an additional shift symmetry
\begin{equation}
\phi \to \phi + c \ ,
\end{equation}

where $c$ is a constant vector.\footnote{For a compact $\phi$ in three dimensions the symmetry can be interpreted as a topological $U(1)$ that acts on the corresponding magnetic monopoles $e^{i \phi}$. For bosonic strings on a worldsheet this symmetry corresponds to space-time translations. } In the absence of boundary potential we can choose Neumann b.c.'s, which will preserve both of these symmetries.

\quad The boundary potential will break the bulk shift symmetry, but we will assume that it preserves the $O(N)$ symmetry. Equation of motion together with the boundary condition describing the system in \eqref{our model} read
\begin{equation} \label{Higg EoM}
\begin{aligned}
\p^2\phi &= 0 \ , \quad \p_\perp\phi |_{x_\perp=0} =  V'(\phi)|_{x_\perp=0} \ .
\end{aligned}
\end{equation}

If the potential has any non-trivial minima these equations admit a constant solution $\phi= \vev{\phi}\neq 0$ satisfying 
\begin{equation}
\frac{\p V}{\p\ph^i}(\vev{\phi}) = 0 \; .
\end{equation}

We will furthermore assume that the solution is a stable minimum with $\frac{\p^2V}{\p\ph^i\p\ph^j}\geq 0$ (by this we mean that the Hessian matrix has only non-negative eigenvalues). Now what are the consequence of having such solution? The vacuum $\vev{\phi}$ will break the global symmetry $O(N)$ down to $O(N-1)$. Had there been no boundary interaction this would obviously not be the case since the new vacuum would be related to the trivial one by the shift symmetry. We will now demonstrate that in the presence of a boundary the expansion around $\vev{\phi}$ leads to a distinct qualitative picture.
 
\quad By running the usual textbook arguments leading to the Goldstone theorem we see that the matrix $\frac{\p^2V}{\p\ph^i\p\ph^j}$ has exactly $N-1$ vanishing eigenvalues corresponding to the broken generators of $O(N)$. We can choose the usual parametrisation to expand about the minimum
\begin{equation} \label{Vev Exp}
\begin{aligned}
\phi = e^{\eta^kT^k}(\vev{\ph} + \s)\; .
\end{aligned}
\end{equation}

Here $T^{k} \ , k\in\{1, ..., N - 1\}$ is the generator of the Lie algebra corresponding to $O(N)/O(N - 1)$ and $\sigma$ is a vector in the flavour space parallel to $\vev{\phi}$ satisfying $|\sigma| \ll |\vev{\ph}|$. If we insert \eqref{Vev Exp} into the potential \eqref{our model} we find that $\eta^k$ is a free massless field and that $\sigma$ has a positive boundary mass and both cubic and quartic interactions\footnote{Here we used that $e^{\eta^kT^k} \in O(N)/O(N - 1) \subset O(N)$, which means that $\ph^2 = (\vev{\ph} + \s)^2$.}
\begin{equation}
\begin{aligned}
S[\eta, \s] &= \int_{\mathbb{R}^d_+}d^dx \left[ \frac{(\p_\m\eta^k)^2}{2} + \frac{(\p_\m\s)^2}{2} + \de(x_\perp) V(\s) \right ]+ \dots \ , \\
V(\eta, \s) &=   \frac{m}{2}\s^2 + \mathcal{O}(\s^3) \ ,
\end{aligned}
\end{equation}

where $m>0$ corresponds to the nonzero eigenvalue of $\frac{\p^2V}{\p\ph^i\p\ph^j}$. This mass term induces a boundary RG flow for $\sigma$ into Dirichlet boundary condition in the IR (by IR we mean large distances parallel to the boundary). The fields $\eta^k$ are similar to the usual Goldstone bosons in that they gain no boundary potential and therefore will retain the Neumann b.c.'s in the IR. This gives us a clear picture of how the symmetry breaking is realized in the IR: the flow will leave us with $N-1$ free Neumann scalars preserving $O(N - 1)$- and the shift-symmetry. The remaining field satisfies Dirichlet condition and therefore its boundary propagator vanishes. This is similar to the tachyon condensation in open string theory \cite{Harvey:2000na} with the preserved $O(N - 1)$- and shift-symmetry being the rotations and translations preserving the IR D-brane. 

\quad In a quantum theory the constant solution to \eqref{Higg EoM} can only exist in the absence of bulk couplings. Were there any bulk couplings the solution to equations of motion would acquire a dependence on the normal coordinate and we would need to deal with the renormalisation of $\phi$ in the near boundary limit.\footnote{By this we mean that the field enjoys the boundary operator expansion $\phi= x_\perp^{-\Delta+\hat{\Delta}} \hat{\phi}+ \dots$, where $\hat{\phi}$ is a boundary operator of dimension $\hat{\Delta}$. As shown in \cite{Prochazka:2019fah}, this expansion is actually equivalent to operator renormalisation and $\hat{\phi}$ can be interpreted as renormalised field.} As a consequence the v.e.v.'s of bulk and boundary fields become unequal, which leads to so called extraordinary phase transitions (see \cite{Diehl:1996kd} for a comprehensive review of phase transitions with boundaries).

\quad In the case of a free bulk that we consider here, the v.e.v. of a bulk field $\phi$ is completely determined from the boundary potential $V$. This is in line with the fact that in absence of bulk interactions, $\phi$ does not renormalise at the boundary (i.e. $\lim_{x_\perp \to 0} \phi = \hat{\phi}$ is well defined).\footnote{See \cite{Prochazka:2019fah, Giombi:2019enr} and the earlier work  \cite{PhysRevB.44.6642} for a proof of this statement.} Thus to understand the IR dynamics of such fields theories we need to determine the potential at the quantum level.

\quad For a potential without non-zero local minima we have two possibilities. Either there exists a boundary RG flow into an IR FP satisfying conformal b.c.'s\footnote{The conformal b.c.'s of \cite{Cardy:1984bb} imply vanishing of the normal-parallel components of the  bulk energy-momentum at the boundary. It was shown in \cite{Prochazka:2019fah} that for models of the kind \eqref{our model} this is equivalent to vanishing of the boundary beta functions.} or new minima appear through quantum corrections. The former scenario is analogous to second-order phase transitions in statistical physics as it involves an IR FP with calculable critical exponents (scaling dimensions of boundary operators). The latter corresponds to a fluctuation induced first-order phase transition with the order parameter $\vev{\phi}$. At the perturbative level the quantum corrections to the classical potential come from the loops through the Coleman-Weinberg (CW) mechanism \cite{PhysRevD.7.1888}. In section \ref{Sec:Tr} we will show how to compute them at the one-loop level for theories of type \eqref{our model}. We illustrate how these ideas can be implemented in a scalar theory with $O(N) \oplus O(N)$-symmetry with interactions confined to the boundary in section \ref{Sec: Dual scalar}. Finally in section \ref{Sec: Disc}, we discuss the broader picture and some future extensions of this work.

\section{One loop effective boundary potentials} \label{Sec:Tr}

In the following we will assume the existence of a classical potential $V(\phi)$ at the boundary. For simplicity we will consider a single scalar field in the bulk, and later generalize this to $O(N)$. We will expand the action \eqref{our model} with $\phi= \phi_{\text{cl}}+ \delta \phi$ about the classical minimum background $\phi_{\text{cl}}$ satisfying the equations of motion \eqref{Higg EoM}.\footnote{There is a factor of $\hbar = 1$ in front of the quantum fluctuations $\de\ph$ and $\de\ch$.} The linear terms vanish by virtue of the equations of motion and we will only keep the quadratic part of the potential
\begin{equation}
V_{quad}=  \frac{M}{2} (\delta \phi)^2 + \mathcal{O}\left(\delta \phi^3 \right) \; , \quad M=  V''(\phi=\phi_{\text{cl}})>0 \; .
\end{equation}

The bulk action for $\delta \phi$ will be the one of a free massless scalar. The one loop effective potential will therefore be obtained by computing the functional determinant of the operator
\begin{equation}
D= - \partial^2 \ ,
\end{equation}

subject to the following boundary condition 
\begin{equation} \label{eq:Mbc}
\partial_\perp \phi |_{x_\perp=0} = M \phi |_{x_\perp=0} \; .
\end{equation}

In general a functional determinant of a differential operator $D$ is computed using the formula
\begin{equation}
\det{D}= e^{-\frac{1}{2}\tr\log{D}} \; ,
\end{equation}

where the trace is evaluated in a suitable (complete) basis of functions $\{ \phi_n \}$. I.e. we have
\begin{equation}
\tr\log{D}= \sum_{n} \int_{\mbr^d_+} d^d x \phi_n^* \log{D} \phi_n \; .
\end{equation}

Without a boundary one typically takes the complete set of eigenfunctions of $D$. For example in the case of $D= - \partial^2$ we take $\phi_n \to \phi_p= e^{i px}$ and the sum over $n$ turns into a momentum space integral.

\quad In our case we have to impose the boundary condition \eqref{eq:Mbc} on the eigenfunctions. The corresponding functional determinant will take the form
\begin{equation}
\tr\log{D}= \int_{\mathbb{R}^d}\frac{d^dp}{(2 \pi)^d} \int_{\mathbb{R}^d_+} d^d x \tilde{\phi}_p^* \log{D} \tilde{\phi}_p \; ,
\end{equation}

with the momentum eigenfunctions satisfying \eqref{eq:Mbc}. More concretely they read
\begin{equation} \label{eq:FdetTilde}
\tilde{\phi}_p(x)= \frac{1}{\sqrt{2}}\left(e^{ipx}+ \frac{i p_\perp-M}{i p_\perp + M} e^{i \tilde{p} x} \right) \; ,
\end{equation}

where we defined a reflected momentum $\tilde{p} = (p_\parallel , - p_\perp)$. By substituting these eigenfunctions in \eqref{eq:FdetTilde} we get
\begin{equation}\label{eq:FdetTildeExplicit}
\tr\log{D} = \int_{\mathbb{R}^d}\frac{d^dp}{(2 \pi)^d}\int_{\mathbb{R}^d_+} d^d x  \left(1  -i\frac{-M+ i p_\perp}{p_\perp- iM} e^{-2 i p_\perp x_\perp}   \right) \log(p^2) \; .
\end{equation}

The first term inside the bracket in \eqref{eq:FdetTildeExplicit} corresponds to the usual (IR divergent) bulk contribution. The second term is a new boundary contribution. We can evaluate it by first calculating the integral over $p_\perp$
\begin{equation} \label{eq:PperpInt}
\int_\mbr dp_\perp \left(-i\frac{M+ i p_\perp}{p_\perp+ iM} \right) [\log(|p_\parallel|+ i p_\perp)+ \log(|p_\parallel|- i p_\perp)]e^{-2 i p_\perp x_\perp} \ .
\end{equation}

\begin{figure} 
	\centering
	\includegraphics[width=0.5\textwidth]{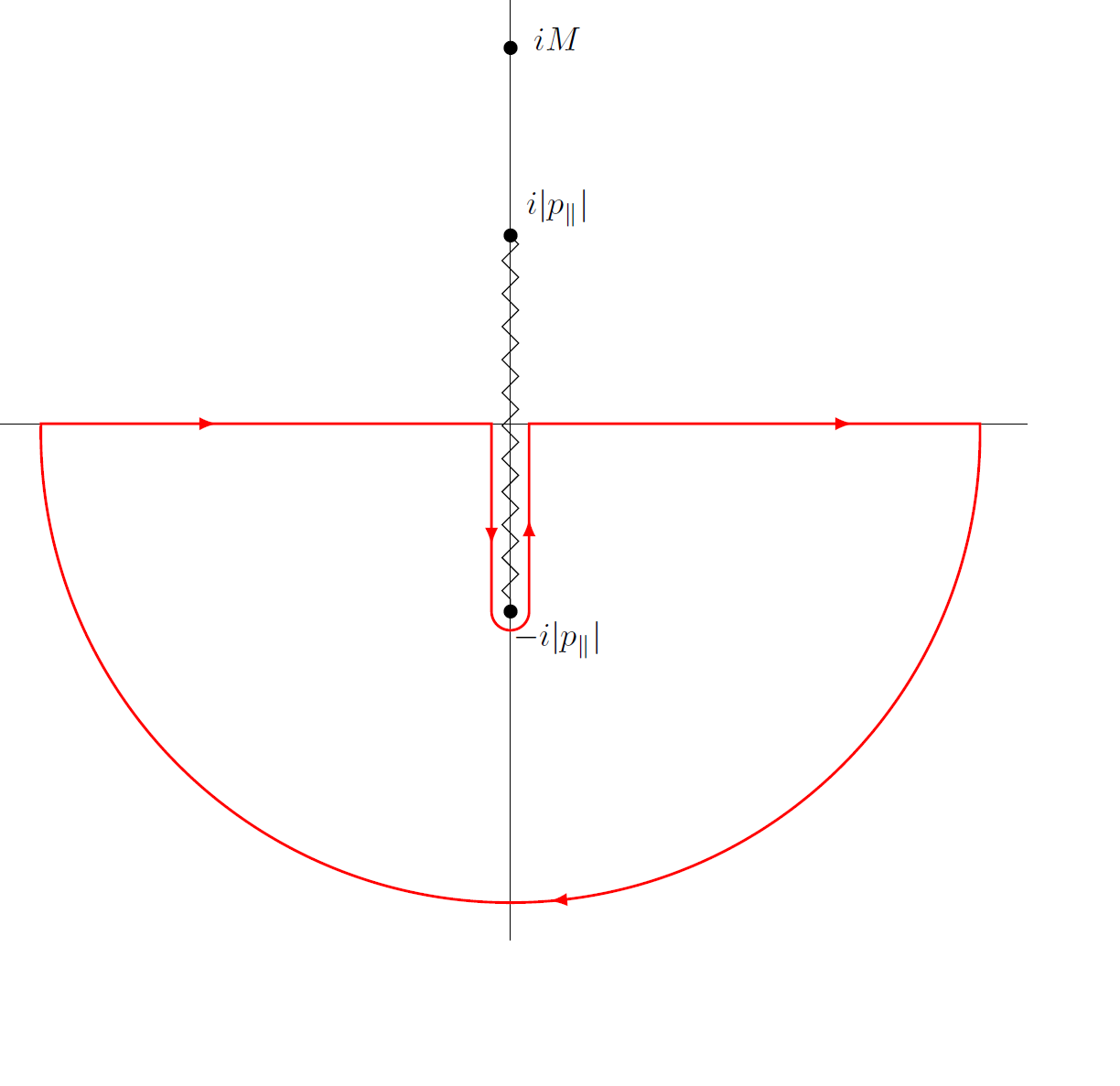}
	\caption{Integration countour for $M>|p_\parallel|$ closed in the lower plane with a branch cut between $(-i |p_\parallel|, +i |p_\parallel|)$ and a simple pole at $iM$. In the case $M<|p_\parallel|$ we take the branch cuts between infinity and $\pm i |p_\parallel|$, which will yield the same result.}
	\label{Fig:Countour}
\end{figure}

This integral is evaluated by using the contour shown on figure \ref{Fig:Countour}. We close the contour in the lower half-plane so that the integral along the semicircle at infinity vanishes. This will also imply that the residue at $i M$ will not contribute. The integral \eqref{eq:PperpInt} will therefore reduce to integrating the segment  around the branch point at $-i |p_\parallel|$ which evaluates to
\begin{equation} \label{eq:PperpInt1}
2\pi \int_{0}^{|p_\parallel|} du \frac{u-M}{u+M} e^{- 2 x_\perp u}\; .
\end{equation}

This expression is still to be integrated over $x_\perp > 0$, which will turn \eqref{eq:PperpInt1} into \eqref{eq:PperpInt1} into
\begin{equation}\label{eq:PperpInt2}
2\pi \frac{1}{2}\int_{0}^{|p_\parallel|} du \frac{u-M}{u(u+M)}   \; .
\end{equation}

The integral in the above expression can now be evaluated by standard methods
\begin{equation}
\int_{0}^{|p_\parallel|} du \frac{u-M}{u(u+M)}= -\log{\left(\frac{|p_\parallel|}{\mu_{\text{IR}}} \right)}+ 2\log{\left(\frac{|p_\parallel|+M}{M} \right)} \; ,
\end{equation}

where $\mu_\text{IR}$ is an IR cutoff introduced to regulate the IR divergence in the above integral.\footnote{Physically this divergence arises from the infinite volume limit (or more specifically it comes from the $x_\perp \to \infty$ region of the original integral).} Finally putting everything together we find the boundary contribution to the functional determinant \eqref{eq:FdetTildeExplicit}
\begin{equation}\label{eq:FdetBoundary}
\tr\log{D}|_{\partial} = \int_{\mbr^{d - 1}} d^2x_{\parallel} \int_{\mbr^{d - 1}} \frac{d^{d-1} p_\parallel}{(2\pi)^{d - 1}}  \left[ -\frac{1}{2} \log{\left(\frac{|p_\parallel|}{\mu_{\text{IR}}} \right)} + \log{\left(\frac{|p_\parallel|+M}{M} \right)} \right] \; .
\end{equation}

The first term in \eqref{eq:FdetBoundary} does not depend on $M$ and therefore will not contribute to the effective potential. So we are left with the second term. From the path integral we have that
\begin{equation}
-\frac{1}{2}\tr\log{D}= V_{\text{eff}}^{\text{1-loop}} + \dots \; ,
\end{equation}

where the dots stand for derivative corrections.
Thus we find that the non-trivial contribution to the boundary effective potential at one loop reads
\begin{equation} \label{eq:1lEff1}
-\int_{\mbr^{d - 1}} \frac{d^{d-1} p_\parallel}{(2 \pi)^{d - 1}}   \log{\left(\frac{|p_\parallel|+M}{M} \right)} \; .
\end{equation}

Note that the numerator of the logarithm in \eqref{eq:1lEff1} leads to a non-analytic power divergence $\Lambda^{d-1} \log M$. Such term does not appear in the usual bulk CW computation, but we can choose a suitable subtraction scheme to remove it  so the relevant 1-loop contribution to the effective potential reads
\begin{equation} \label{eq:1lEff2}
V_{\text{eff}}^{\text{1-loop}}=  -\int_{\mbr^{d - 1}} \frac{d^{d-1} p_\parallel}{(2 \pi)^{d - 1}}   \log{\left(|p_\parallel|+M\right)} \; .
\end{equation}

Note that for $N>1$ this formula still holds with $M$ promoted to a matrix.

\section{$O(N_\ph)\oplus (N_\ch)$ scalar model} \label{Sec: Dual scalar}

\subsection{The model}

In this section we will consider an $O(N_\ph)\oplus (N_\ch)$ scalar model similar to that in \cite{PhysRevB.13.412, Calabrese:2002bm, Beekman:2019sof}, but with interactions happening at the boundary instead of in the bulk. The model will be defined by the following action 
\begin{equation} \label{s-Action}
\begin{aligned}
S[\ph, \ch] &= \int_{\mathbb{R}^d_+}d^dx \left( \frac{(\p\ph)^2}{2} + \frac{(\p\ch)^2}{2} + \de(x_\perp) V(\ph^2, \ch^2) \right) \ , \\ 
V(\ph^2, \ch^2) &= \frac{\la}{8}\ph^4 + \frac{\x}{8}\ch^4 + \frac{g}{4}\ph^2\ch^2 \ .
\end{aligned}
\end{equation}

The scalar fields $\ph \equiv \ph^i \ , i\in\{1, ..., N_\ph\}$ and $\ch \equiv \ch^a \ , a\in\{1, ..., N_\ch\}$ satisfy $O(N_\ph)$- and $O(N_\ch)$-symmetry respectively. The mixed interaction breaks the $O(N_\ph + N_\ch)$-bulk symmetry down to $O(N_\ph)\oplus O(N_\ch)$.

\quad To simplify the computations we take $N_\ph = N_\chi \equiv N$ and therefore also $\lambda= \xi$. In which case the theory also has an additional $\mathbb{Z}_2$ symmetry
\begin{equation} \label{Central Symm}
\begin{aligned}
\ph\longleftrightarrow \ch \ .
\end{aligned}
\end{equation}

From dimensional analysis we have the following engineering dimensions
\begin{equation} \label{Dim}
\begin{aligned}
\D_\ph = \D_\ch &= \frac{d - 2}{2} = \frac{1 - \e}{2} \ , \quad \D_\la = \D_\xi = \D_g = 3 - d = \e \ .
\end{aligned}
\end{equation}

A detailed discussion of the renormalisation of such models has been presented in our earlier work \cite{Prochazka:2019fah}. In appendix \ref{App: Scalar beta} we compute the $\be$-functions for a model with generic $\lambda, \xi$ up to order two in the coupling constants. For $\xi=\lambda$ we have the following beta functions
\begin{equation} \label{s-Beta fcns}
\begin{aligned}
\be_\la &= - \e\la + \frac{N + 8}{4\pi}\la^2 + \frac{N}{4\pi}g^2 + ... \ , \quad \be_g = - \e g + \frac{g^2}{\pi} + 2\frac{N + 2}{4\pi}\la g  + ... \ .
\end{aligned}
\end{equation}

\begin{figure} 
	\centering
	\includegraphics[width=0.5\textwidth]{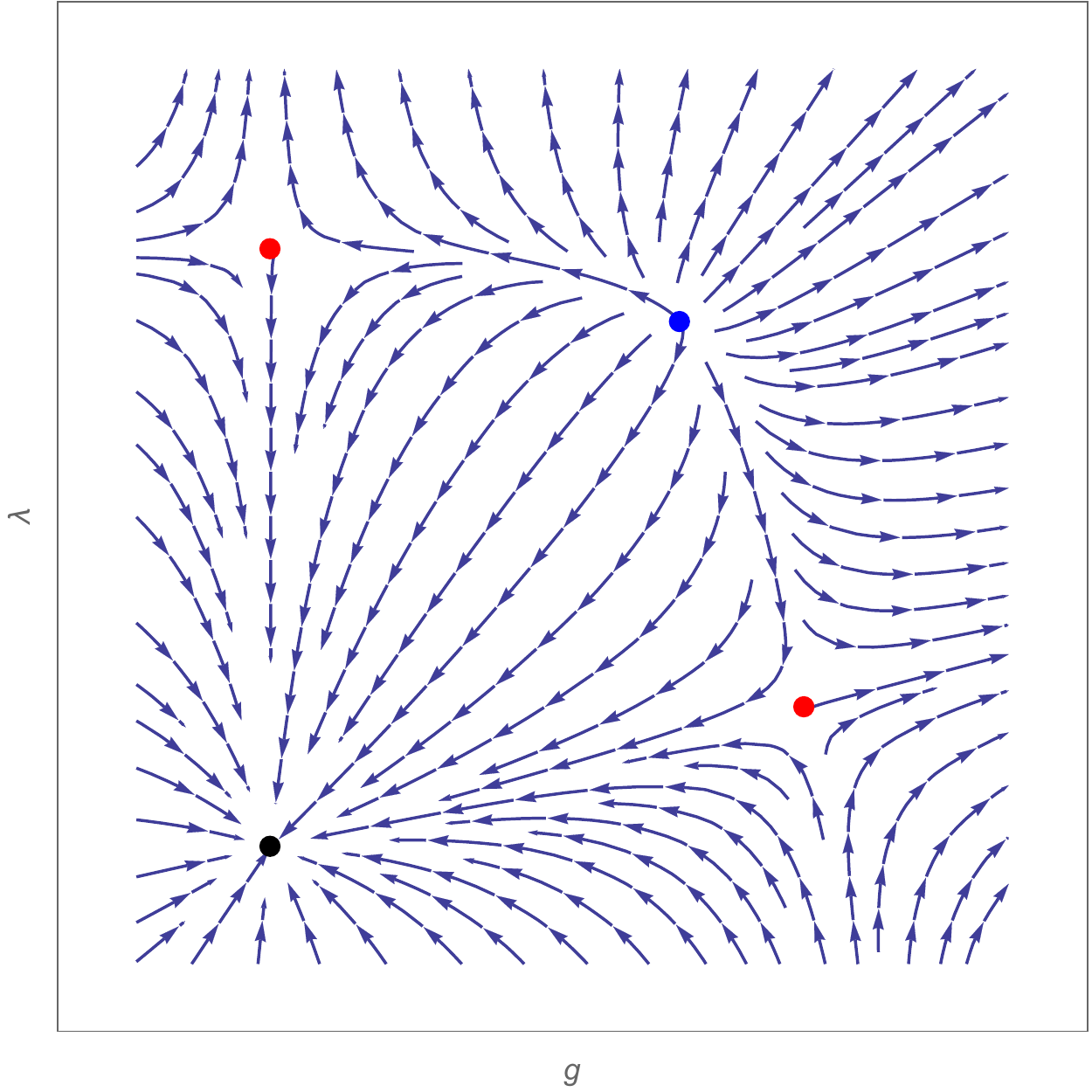}
	\caption{The RG flow for the model \eqref{s-Action} when $N_\ph = N_\ch = 1$. FP's are marked by dots, where the black dot is the fully repellent Gaussian FP (the physical RG flow goes in the opposite direction), the red dots define a separatrix that separates regions corresponding to first- and second-order phase transitions and the blue dot is a fully attractive FP that is stable in the IR.} 
	\label{Fig: RGFlow}
\end{figure}

These $\be$-functions have one Gaussian, and three WF FP's defining a boundary RG flow chart depicted on figure \ref{Fig: RGFlow}. The positions of these FP's read
\begin{equation} \label{eq:FPs}
\begin{aligned}
(g^*, \la^*) &\in \left\{ (0, 0), \left( 0, \frac{4\pi}{N + 8}\e \right), \left( \frac{2 \pi  (4-N )}{N^2+8} \e, \frac{2 \pi  N}{N^2+8} \e\right), \left( \frac{2 \pi}{N+4} \e, \frac{2 \pi }{N+4} \e\right) \right\} \ .
\end{aligned}
\end{equation}

The first one is the fully repulsive Gaussian FP, the second of these corresponds to decoupled $O(N)$ models with a single coupling at a WF point studied in \cite{Prochazka:2019fah, Giombi:2019enr}, the third one defines a stable solution only for $N<4$, while the last FP enjoys an emergent $O(2N)$ symmetry. As already mentioned, the fundamental field $\phi$ does not acquire an anomalous dimension at these FP's. On the other hand the composite operators (eg. $\ph^2, \ch^2$ etc.) have to be renormalised due to divergences in the boundary limit which leads to their anomalous dimensions in perturbation theory \cite{Prochazka:2019fah}.

\quad The flow diagram in figure \ref{Fig: RGFlow} shares many features to the corresponding charts of the Abelian-Higgs model or the bulk $O(N)\oplus O(N)$ model (see for example \cite{Beekman:2019sof}). In particular the diamond region corresponds to the domain of attraction of the symmetric, IR stable critical point. One would expect that the separatrix running from the Gaussian FP to the third FP in \eqref{eq:FPs} (which is similar to tri-critical FP in the language of statistical physics) should determine the cross-over to a region with fluctuation-induced first order phase transition. More specifically the RG flow in this region should end up in an ordered phase. In the next section we will argue that this is indeed the case. \\

\subsection{Coleman-Weinberg mechanism} \label{Sec: Dual scalar CW}
In this section we will follow the standard reasoning of Coleman and Weinberg \cite{PhysRevD.7.1888} applied within the context of this paper. We will expand around classical field values
\begin{equation} \label{Classical scalars}
\begin{aligned}
\ph^i &= \ph^i_{cl} + \de\ph^i \ , \quad \left|\de\ph^i\right| \ll 1 \ , \\
\ch^a &= \ch^a_{cl} + \de\ch^a \ , \quad \left|\de\ch^a\right| \ll 1 \ .
\end{aligned}
\end{equation}

We will only keep up to quadratic terms
\begin{equation} \hspace{-10px}
S = S[\ph_{cl}, \ch_{cl}] + \int_{\mbr^d_+}d^dx \left( \frac{(\p\de\ph)^2}{2} + \frac{(\p\de\ch)^2}{2} + \de(x_\perp) \de V(\ph_{cl}^2, \de\ph^2, \ch_{cl}^2, \de\ch^2) \right) \ .
\end{equation}
where the quadratic part of the potential can be written as a boundary mass term
\begin{equation}
\de V = \Ph^I\left(m_\Ph^{IJ}\right)\Ph^J \ , 
\end{equation}
with
\begin{equation}
\begin{aligned}
m_\Ph^{IJ} &= \begin{pmatrix}
m_\ph^{IJ} & g\ph_{cl}^j\ch_{cl}^b \\
g\ch_{cl}^a\ph_{cl}^k & m_\ch^{ab}
\end{pmatrix}^{IJ} \ ,
\end{aligned}
\end{equation}
\begin{equation} \label{dual scalar masses}
\begin{aligned}
m_\ph^{ij} &\equiv A^\la_g\de^{ij} + \la\hp_{cl}^i\hp_{cl}^j \ , \\ 
m_\ch^{ab} &\equiv A^g_\la\de^{ab} + \la\hc_{cl}^a\hc_{cl}^b \ , \\
A(x, y) &= \frac{x\ph_{cl}^2 + y\ch_{cl}^2}{2} \ .
\end{aligned}
\end{equation}
Here we defined the field $\Ph^I = \de\ph^j\oplus\de\ch^a \ , I \in \{1, ..., N, N + 1, ..., 2N\}$. The one loop correction to the path integral $Z_\Ph$ can be calculated by substituting the above mass term into the formula derived in section \ref{Sec:Tr} and performing the relevant momentum integral \eqref{eq:1lEff2}. We leave the details of this computation to appendix \ref{App: Scalar det}. It yields the effective boundary potential
\begin{equation} \label{s-Eff Pot} 
\begin{aligned}
V_\eff(\ph_{cl}^2, \ch_{cl}^2) &= \lim\limits_{\La\rightarrow\infty} \left[ \left( \frac{\la}{8} + B_1 \right)\ph_{cl}^4 + \left( \frac{g}{4} + B_2 \right) \ph_{cl}^2\ch_{cl}^2 + \left( \frac{\x}{8} + B_3 \right) \ch_{cl}^4 + \rig \\
\eq\lef + \X_1 + \X_2 + A_1\ph_{cl}^2 + A_2\ch_{cl}^2 \frac{}{} \right] \ .
\end{aligned}
\end{equation}

Here $\X_1 \ , \X_2$ can be found in appendix \ref{App: Scalar det}, and the constants $A_i, B_j \ , i\in\{1,2\} \ , j\in\{1,2,3\}$ are counter-terms (which depend on the momentum cut-off $\La \gg 1$) which can be fixed by defining the renormalised masses and coupling constants as the respective coefficients in the effective potential
\begin{equation}
\begin{aligned}
\left.\frac{\partial V}{\partial(\ph_{cl}^2)}\right|_{\ph_{cl}^2 =\ch_{cl}^2 = 0} = \left.\frac{\partial V}{\partial(\ch_{cl}^2)}\right|_{\ph_{cl}^2 =\ch_{cl}^2 = 0} &= 0 \ , \\
\left.\frac{\partial^2V}{\partial(\ph_{cl}^2)^2}\right|_{\ph_{cl}^2 =\ch_{cl}^2 = 0} = \left.\frac{\partial^2V}{\partial(\ch_{cl}^2)^2}\right|_{\ph_{cl}^2 =\ch_{cl}^2 = 0} &= \frac{\la}{4} \ , \\
\left.\frac{\partial^2V}{\partial(\ph_{cl}^2)\partial(\ch_{cl}^2)}\right|_{\ph_{cl}^2 =\ch_{cl}^2 = 0} &= \frac{g}{4} \ .
\end{aligned}
\end{equation}

The latter two derivatives are IR divergent in the $\phi_{cl}, \chi_{cl} \to 0$ limit due to the presence of logarithms in $V_{\eff}$. Following the CW procedure we can resolve this issue by evaluating the renormalisation conditions at non-zero field value for $\ph$ (alternately for $\ch$)
\begin{equation} \label{Ren cond}
\begin{aligned}
\left.\frac{\partial V_\eff}{\partial(\ph_{cl}^2)}\right|_{\ph_{cl}^2 =  \ch_{cl}^2 = 0} = \left.\frac{ \partial V_\eff}{\partial(\ch_{cl}^2)}\right|_{\ph_{cl}^2 = \ch_{cl}^2 = 0} &= 0 \ , \\
\left.\frac{\partial^2V_\eff}{\partial(\ph_{cl}^2)^2}\right|_{\ph_{cl}^2 = \m, \ch_{cl}^2 = 0} = \left.\frac{\partial^2V_\eff}{\partial(\ch_{cl}^2)^2}\right|_{\ph_{cl}^2 = \m, \ch_{cl}^2 = 0} &= \frac{\la}{4} \ , \\
\left.\frac{\partial^2V_\eff}{\partial(\ph_{cl}^2)\partial(\ch_{cl}^2)}\right|_{\ph_{cl}^2 = \m, \ch_{cl}^2 = 0} &= \frac{g}{4} \ ,
\end{aligned}
\end{equation}
where $\mu$ is an arbitrary RG scale and we used that near $d=3$ the scaling dimension of $\phi_c$ is \eqref{Dim} so to leading order in $\epsilon$-expansion $\ph_{cl}^2$ scales as mass.\footnote{Note that we choose to define the renormalisation conditions w.r.t. $\ph_{cl}^2, \chi_{cl}^2$ as opposed to some particular component of $\phi_{cl}, \chi_{cl}$. In this way we obtain $O(N)$ invariant counter-terms, but otherwise the physics remains the same.}  

\quad The renormalisation conditions \eqref{Ren cond} now fix the counter-terms in such a way that the divergences in $\La$ vanish in the effective potential
\begin{equation}
\begin{aligned}
A_1 &= (d - 4)e^{(d - 3)\g_E/2}\frac{N g + (N + 2)\la}{2^d\pi^{(d - 1)/2}}\La^{d - 2} \ , \\ 
A_2 &= (d - 4)e^{(d - 3)\g_E/2}\frac{N g + (N + 2)\x}{2^d\pi^{(d - 1)/2}}\La^{d - 2} \ .
\end{aligned}
\end{equation}
\begin{equation}
\begin{aligned}
B_1 &= \left( \frac{N + 8}{4\pi}\la^2 + \frac{N}{4\pi}g^2 \right) \frac{\log\La}{8} + ... \ , \\
B_2 &= \left( \frac{g^2}{\pi} +2 \frac{N + 2}{4\pi}\la g  \right) \frac{\log\La}{4} + ... \ .
\end{aligned}
\end{equation}

Here we only wrote out the divergent parts of the bare couplings \eqref{Bare couplings} in $B_j \ , j\in\{1, 2\}$. As a consistency check we can readily verify that the coefficients of the logarithmic divergences in $B_j$ agree with the beta functions \eqref{s-Beta fcns} computed with dimensional regularisation. If we plug these constants into \eqref{s-Eff Pot} we get the full effective potential which we do not write out here since it is given by a cumbersome expression. A details of this can be found in an appended Mathematica notebook.

\quad One can verify by explicit computation that this effective potential admits a perturbative minimum at $\ph_{cl}^2 = \mu$ with
\begin{equation} \label{Minimum eq}
\begin{aligned}
\ph_{cl}^i\left.\frac{\partial V}{\partial\hp_{cl}^i}\right|_{\ph_{cl}^2 = \vev{\ph}^2= \m, \ch_{cl}^2 = \vev{\ch}^2= 0} &= \ch_{cl}^a\left.\frac{\partial V}{\partial\ch_{cl}^a}\right|_{\ph_{cl}^2 = \vev{\ph}^2= \m, \ch_{cl}^2 = \vev{\ch}^2=0} &= 0 \ ,
\end{aligned}
\end{equation}

provided the couplings satisfy the relation
\begin{equation} \label{laCond}
\la = \frac{4\pi - \sqrt{16\pi^2 - 4N(N + 8)g^2}}{2(N + 8)} = \frac{Ng^2}{4\pi} + \mco(g^3) \ .
\end{equation}

\begin{figure}
	\centering
	\includegraphics[width=.40\textwidth]{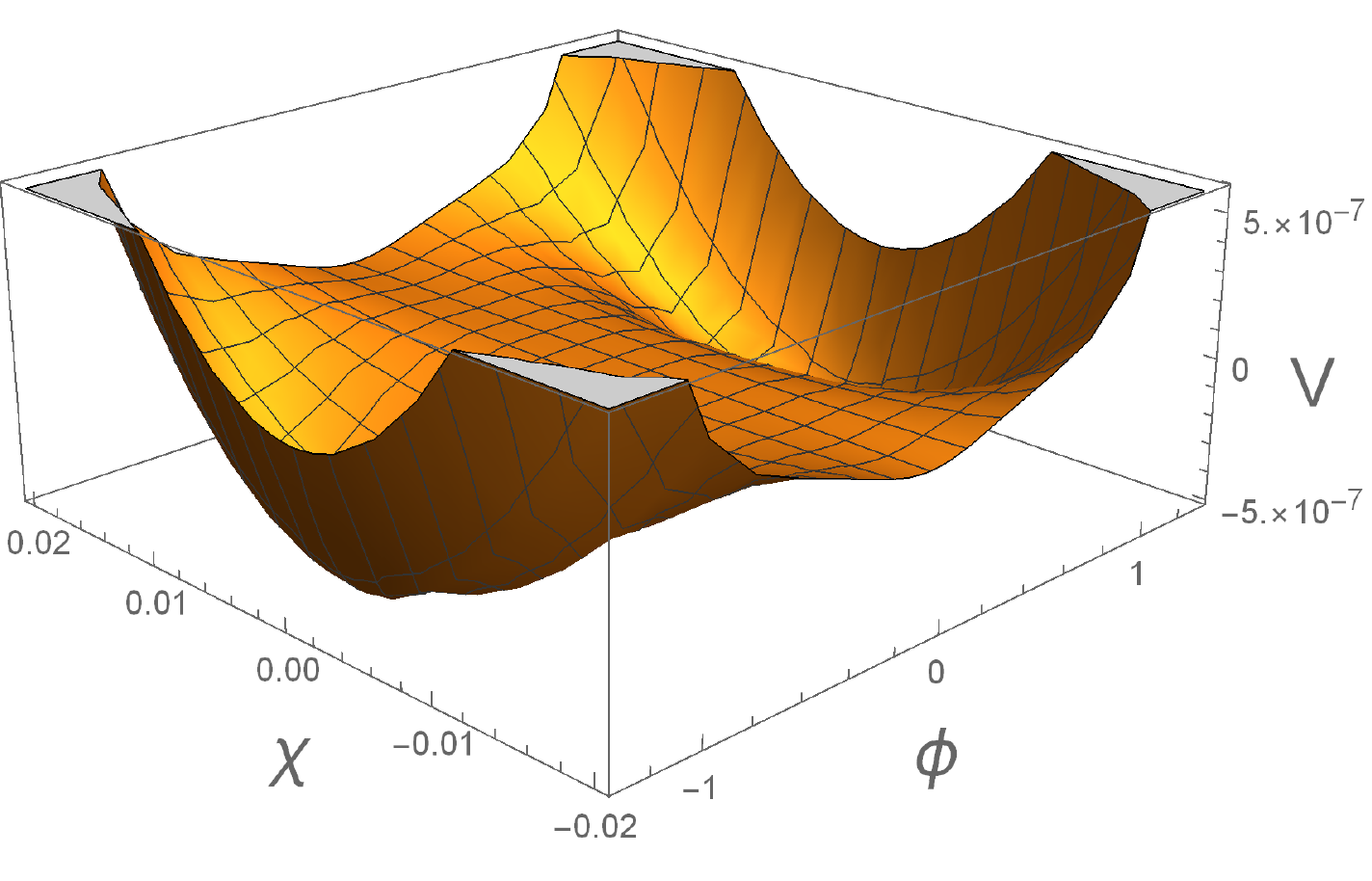} \hspace{30px}
	\includegraphics[width=.40\textwidth]{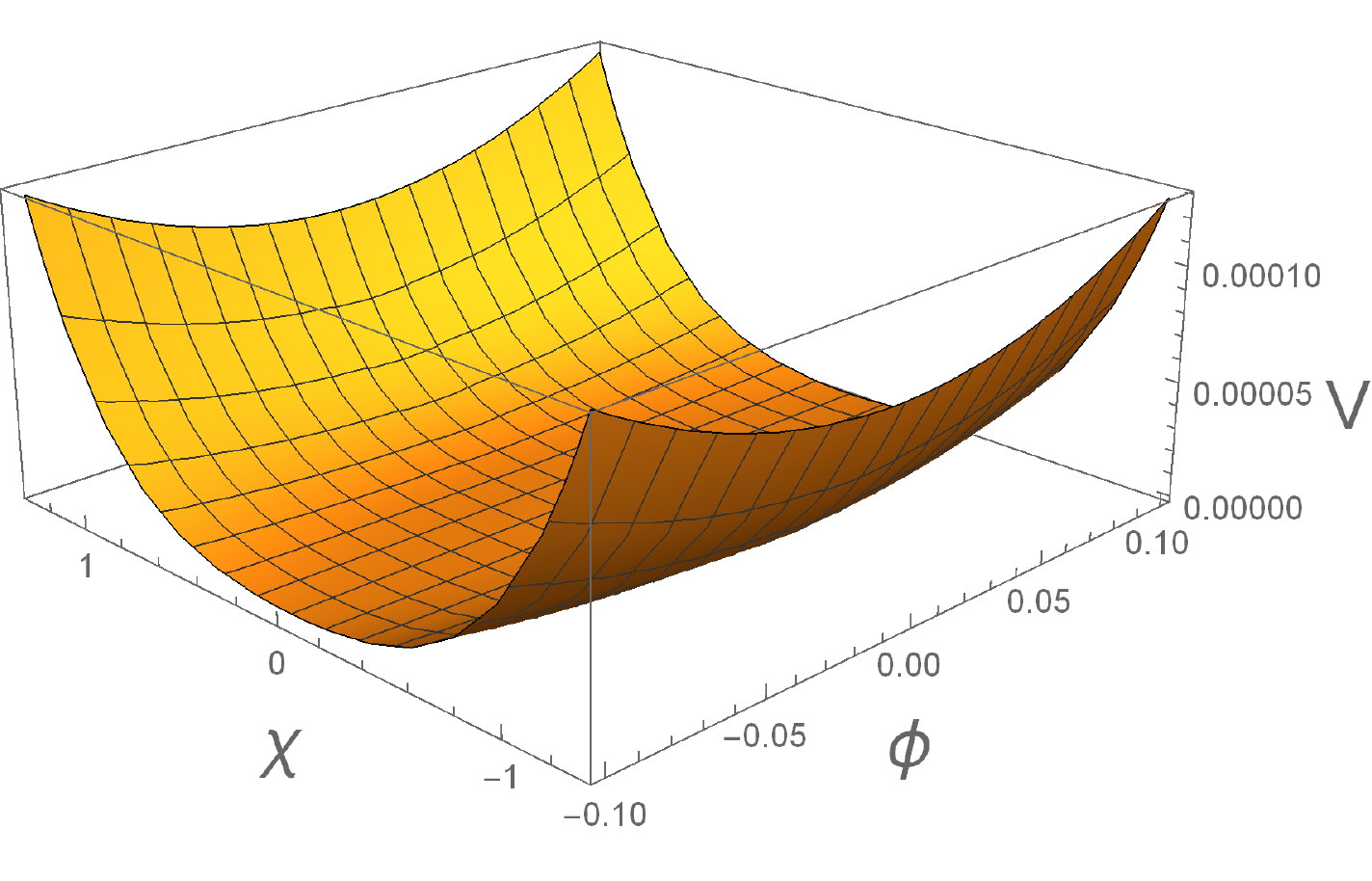}
	\\[\smallskipamount]
	\includegraphics[width=.40\textwidth]{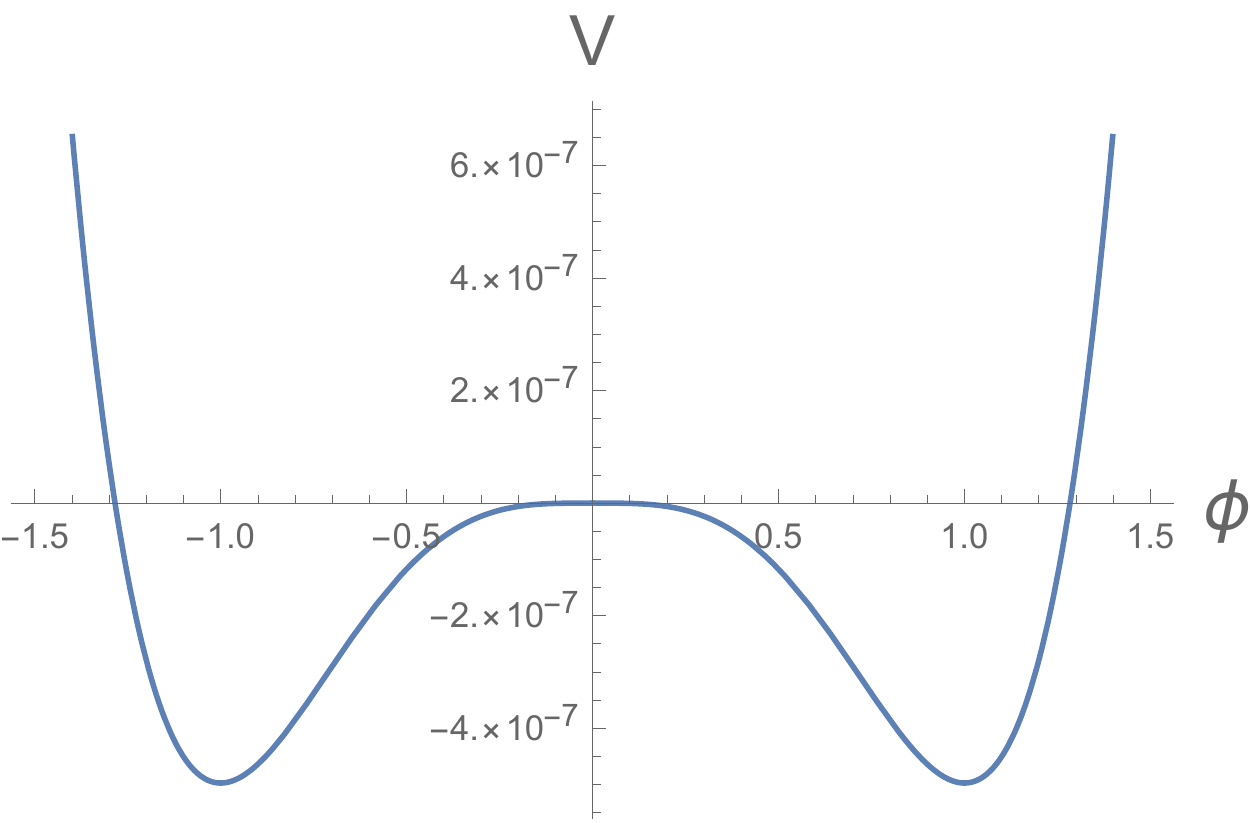} \hspace{30px}
	\includegraphics[width=.40\textwidth]{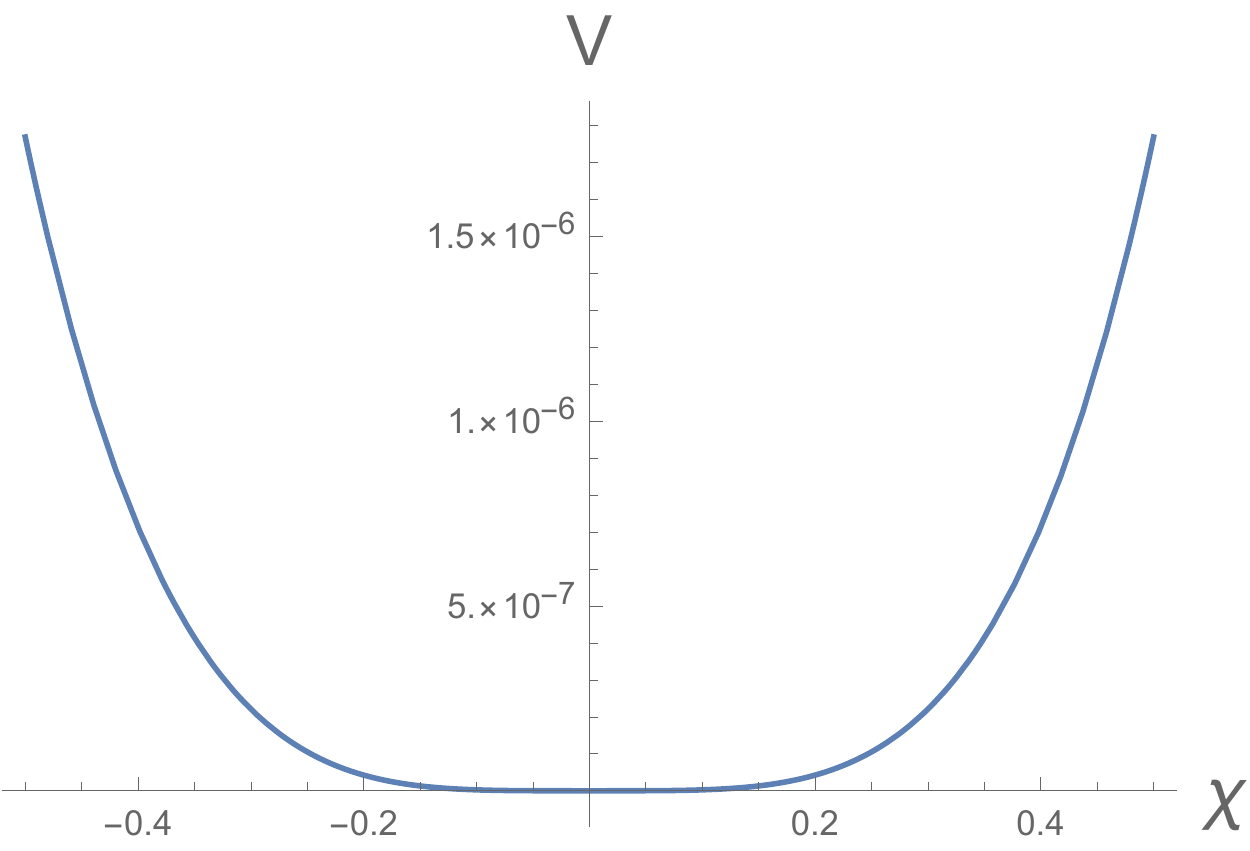}
	\caption{Plots of the effective potential for $N= 1$. There are two three-dimensional plots, one with narrow range of $\ph_{cl}$ and one of $\ch_{cl}$. One can see that the potential only has two minima along the $\ch_{cl}$-axis. The two-dimensional plots are slices of the three-dimensional plot when $\ph_{cl} = 0$ or $\ch_{cl} = 0$. In the plots $g = 0.01$ and $\m = 1$.} \label{Fig: graphs}
\end{figure}

A plot of the effective potential with $N = 1$ is depicted on figure \ref{Fig: graphs} from which we can see that this solution indeed corresponds to a minimum.
Without loss of generality we can parametrise this solution as follows
\begin{equation} \label{VacCW}
\vev{\ph}= (\sqrt{\m},0,... 0) \; , \quad \vev{\ch}= 0 \; .
\end{equation}

This minimum tells us that the $O(N)\oplus O(N)$ symmetry has been broken into $O(N - 1)\oplus O(N)$. Additionally this vacuum breaks the discrete symmetry \eqref{Central Symm}.

\quad Since this vacuum only breaks one of the $O(N)$-symmetries we can apply the arguments discussed in the introduction around \eqref{Higg EoM}. In particular we can now study the perturbations around \eqref{VacCW} by using the parametrisation \eqref{Vev Exp} for $\ph$. Expanding the effective potential to the quadratic order yields\footnote{At higher orders there will be interactions with both even and odd powers of $\s$, e.g. $\s^3$ and $\s\ch_{cl}^2$.}
\begin{equation} \label{CWquadPot}
\begin{aligned}
V_\eff(\s, \ch^2) &= \frac{N g^2\m}{8\pi}\s^2 + \left( 1 - \frac{g}{\pi} \right) \frac{g\m}{4}\ch^2 + \dots  \ ,
\end{aligned}
\end{equation}

where the dots stand for higher order terms in $g, \chi, \sigma$.  The positive sign of both mass terms is a consequence of \eqref{VacCW} being the minimum of the effective potential. The leading (positive) correction to the mass term for $\chi$ is a purely classical consequence of the mixed coupling.  Hence we see that the potential \eqref{CWquadPot} induces a boundary RG flow ending with $N-1$ Neumann scalars from the broken $O(N)$ symmetry.

\quad To summarize, the theory \eqref{s-Action} we started with had $O(N)\oplus O(N)$ symmetry as well as the symmetry \eqref{Central Symm}. After integrating out quantum fluctuations, one of the $O(N)$-symmetries is still preserved while the symmetry \eqref{Central Symm} is completely broken and the other $O(N)$-symmetry is broken down to its subgroup $O(N - 1)$. The remaining $O(N-1)$ can be seen through the effective theory in the IR which contains $N-1$ Neumann scalars (which additionally regain the shift symmetry), and $N + 1$ Dirichlet scalars. 

\quad At last let us discuss the validity of the one-loop approximation and its relevance to the flow diagram charted on figure \ref{Fig: RGFlow}. The condition \eqref{laCond} tells us that the region of validity of the approximation lies in the $\lambda, g>0$ quadrant. Furthermore, in the $g \ll 1$ limit this region lies below the line connecting the Gaussian FP with the 3rd FP in \eqref{eq:FPs}, which is defined by the relation $\lambda= k g$ with $k$ being $O(g^0)$ and positive. As we can see in figure \ref{Fig: RGFlow}, the flow in this region drives the coupling $\lambda$ to negative values and hence we would indeed expect a phase-transition happening here.  We should also remark that the approximation we used cannot be trusted for large (or small) field values far from $\sqrt{\mu}$ and thus we cannot exclude the possibility of other vacua hiding in these regions.

\quad Let us finally mention the $d=3$ case. For $\epsilon=0$ the three non-Gaussian FP's in figure \ref{Fig: RGFlow} disappear and the asymptotic freedom is lost.\footnote{More concretely the boundary RG flow ends with the Gaussian FP with Neumann b.c.'s for all fields. } Despite that, the arguments of this subsection apply if we think of the model at non-zero $(g, \lambda)$ as an effective field theory with radiately generated potential just as in the original Coleman-Weinberg paper.

\section{Discussion} \label{Sec: Disc}

In this paper we have argued that many of the critical phenomena appearing for interacting bulk systems can also be observed in free theories with non-trivial dynamical b.c.'s. These dynamical b.c.'s generically break the conformal symmetry and induce an RG flow at the boundary. We have found that in this context the phase transitions should be understood in terms of the b.c.'s at the IR end of this flow. The second-order phase transitions are described by a boundary RG flow preserving the global symmetries of the theory. It has an IR FP with conformal b.c.'s that are neither Dirichlet nor Neumann. To check whether the FP's we discovered in section \ref{Sec: Dual scalar} are artefacts of the $\e$-expansion or actual physical boundary CFT's would require a non-perturbative approach which is beyond the scope of this paper. An evidence for existence of such FP beyond perturbation theory was nevertheless put forward in a recent work \cite{Behan:2020nsf} employing the numerical boostrap. It would be interesting to investigate the existence of the phase diagram \ref{Fig: RGFlow} by such boostrap methods.

\quad Our analysis also suggests the possibility of RG flows leading to first-order phase transitions induced by quantum effects. These will be described by a combination of Dirichlet and Neumann scalars, with the latter playing a role analogous to Goldstone bosons of the ordinary symmetry breaking. To confirm such assertion beyond the perturbative reasoning offered here, one could devise a lattice simulation of the model.

\quad The physical interpretation of the model described in section \ref{Sec: Dual scalar} remains an open question. It would be very interesting to explore whether the interpolation $\epsilon \to 1$ of the model we described in section \ref{Sec: Dual scalar} describes a meaningful two-dimensional theory. In $d=3$ the free scalar can be interpreted as a dual photon of the Maxwell theory. A boundary potential \eqref{our model} would correspond to a monopole potential breaking the topological $U(1)$ symmetry. Given that the bulk theory is free, it would be very interesting to investigate the possibility of exactly solvable monopole potentials.
 
\quad The free $O(N)$ model with $N>1$ also has a nice condensed matter interpretation as crystaline displacement fields with $N$ being the spatial dimension of the solid \cite{Beekman_2017}. The boundary potential we consider would correspond to dislocations interacting directly at the edge of the solid. It would be amusing to explore whether it can describe a realistic physical situation. 

\quad Let us mention few interesting possible extensions of this work. First we could try coupling the free bulk scalar to boundary degrees of freedom and use this to generate an effective potential and condensates for the boundary fields. This could provide some quantitative arguments for the possible existence of  ordered phases of mixed dimensional theories similar to the ones recently considered in the literature (e.g. \cite{DiPietro:2019hqe, Herzog:2018lqz, Hsiao:2017lch}).

\quad On the other hand we could consider adding bulk couplings and making connection with the recent work \cite{Herzog:2020lel}, where the contribution of a bulk $\phi^6$ interaction to the one-loop effective action was computed.

\newpage

\section*{Acknowledgements}

We are grateful to Hans Werner Diehl, Guido Festuccia for illuminating discussions and to Agnese Bissi for reading the manuscript. VP was supported by the ERC STG grant 639220 during the work on this project and AS is supported by Knut and Alice Wallenberg Foundation KAW 2016.0129.

\appendix

\section{$\beta$-function} \label{App: Scalar beta}

In this appendix we find the $\be$-function for the model \eqref{s-Action}. This is done in the standard quantum field theory way. We will study correlators up to order two in the coupling constants.

\quad In order to find the $\beta$-functions in the model \eqref{s-Action}, we need to study the following correlators
\begin{equation} \label{s-Corr Scalar}
\begin{aligned}
G_{\ph}^{jklm}(p) &= \langle\hp^j(p_1)\hp^k(p_2)\hp^l(p_3)\hp^m(p_4)\rangle \\
&= -\frac{\la_0}{8}8D^{jklm} + \left(-\frac{\la_0}{8}\right)^2 32 \frac{E^{jklm}I_{12} + E^{jlkm}I_{13} + E^{jmkl}I_{14}}{(2\pi)^{d - 1}} + \\
\eq + \left(-\frac{g_0}{4}\right)^28 \frac{\de^{jk}\de^{lm}I_{12} + \de^{jl}\de^{km}I_{13} + \de^{jm}\de^{kl}I_{14}}{(2\pi)^{d - 1}} + ... \ , \\
\end{aligned}
\end{equation}
\begin{equation} \label{s-Corr Fermion}
\begin{aligned}
G_{\ch}^{abcd}(p) &= \langle\hat{\ch}^a(p_1)\hat{\ch}^b(p_2)\hat{\ch}^c(p_3)\hat{\ch}^d(p_4)\rangle \\
&= -\frac{\xi_0}{8}8D^{abcd} + \left(-\frac{\xi_0}{8}\right)^2 32 \frac{E^{abcd}I_{12} + E^{acbd}I_{13} + E^{adbc}I_{14}}{(2\pi)^{d - 1}} + \\
\eq + \left(-\frac{g_0}{4}\right)^28 \frac{\de^{ab}\de^{cd}I_{12} + \de^{ac}\de^{bd}I_{13} + \de^{ad}\de^{bc}I_{14}}{(2\pi)^{d - 1}} + ... \ , \\
\end{aligned}
\end{equation}
\begin{equation} \label{s-Corr Mixed}
\begin{aligned}
G_{\ph\ch}^{jkab}(p) &= \langle\hp^j(p_1)\hp^k(p_2)\hat{\ch}^a(p_3)\hat{\ch}^b(p_4)\rangle \\
&= -\frac{g_0}{4}4\de^{jk}\de^{ab} + \left(-\frac{g_0}{4}\right)^2 16\de^{jk}\de^{ab}\frac{I_{13} + I_{14}}{(2\pi)^{d - 1}} + \\
\eq + \left(-\frac{g_0}{4}\right)\left(-\frac{\la_0}{8}\right)16(N_\ph + 2)\de^{jk}\de^{ab}\frac{I_{12}}{(2\pi)^{d - 1}} + \\
\eq + \left(-\frac{g_0}{4}\right)\left(-\frac{\x_0}{8}\right)16(N_\ch + 2)\de^{jk}\de^{ab}\frac{I_{34}}{(2\pi)^{d - 1}} + ... \ .
\end{aligned}
\end{equation}

Here $\la_0 , g_0$ and $\x_0$ are the bare coupling constants that appear in the action \eqref{s-Action}. Hatted operators denote their respective boundary fields. We have the Wick factors
\begin{equation} \label{Tensor struc}
\begin{aligned}
D^{jklm} &= \de^{jk}\de^{lm} + \de^{jl}\de^{km} + \de^{jm}\de^{kl} \ , \\
E^{jklm} &= (N_\ph + 2)\de^{jk}\de^{lm} + D^{jklm} \ .
\end{aligned}
\end{equation}

The master integral $I_{jk}$ is found using a Julian-Schwinger parametrization and is given by an Euler-Beta function
\begin{equation}
\begin{aligned}
I_{jk} &= I^{d - 1}_{1/2, 1/2}(p_j + p_k) \ , \quad I_{\al\be}^n(p) &= \int_{\mathbb{R}^n}\frac{d^nk}{|p - k|^\al|w|^\be} = \pi^{n/2}\frac{\G_{\al + \be - n/2}}{\G_\al\G_\be}\frac{B_{n/2 - \al, n/2 - \be}}{|p|^{2(\al + \be) - n}} \ ,
\end{aligned}
\end{equation}

where in $d = 3 - \e$ it has a pole in $\e$
\begin{equation*}
\begin{aligned}
I^{d - 1}_{1/2, 1/2}(p) = \frac{1}{2\pi} \left( \frac{1}{\e} + \log\left(\sqrt{\frac{64\pi}{e^{\g_E}}}\right) - \log(p) \right) + \mco(\e) \ .
\end{aligned}
\end{equation*}

The bare coupling constants that renormalises these correlators\footnote{And which absorbs the factors of $\g_E$ and $\log(\pi)$.} are given by
\begin{equation} \label{Bare couplings}
\begin{aligned}
\la_0 &= \left( \frac{e^{\g_E/2}}{64\pi} \right)^{\e/2} \mu^\e \left( \la + \frac{N_\ph + 8}{4\pi}\frac{\la^2}{\e} + \frac{N_\ch}{4\pi}\frac{g^2}{\e} \right) + ... \ , \\
\xi_0 &= \left( \frac{e^{\g_E/2}}{64\pi} \right)^{\e/2} \mu^\e \left( \x + \frac{N_\ch + 8}{4\pi}\frac{\x^2}{\e} + \frac{N_\ph}{4\pi}\frac{g^2}{\e} \right) + ... \ , \\
g_0 &= \left( \frac{e^{\g_E/2}}{64\pi} \right)^{\e/2} \mu^\e \left( g + \frac{g^2}{\pi\e} + \frac{N_\ph + 2}{4\pi}\frac{\la g}{\e} + \frac{N_\ch + 2}{4\pi}\frac{\x g}{\e} \right) + ... \ .
\end{aligned}
\end{equation}

Here the dots represent terms that have more than two coupling constants, $\m$ is the renormalisation scale, and $\la, g$ as well as $\x$ are renormalised coupling constants. Please note that we have used multiplicative renormalisation of $g_0$, and both multiplicative and additive renormalisation of $\la_0$ as well as $\x_0$. We can see that $\la_0$ and $\x_0$ are the same up to flavours. To find the $\be$-functions we will use
\begin{equation} \label{s-Definition Beta}
\begin{aligned}
\frac{\p\log\s_0}{\p\log\m} &= 0 \ , \quad \frac{\p\s}{\p\log\m} = \be_\s \ , \quad \frac{\p\log\s}{\p\log\m} = \frac{\be_\s}{\s} \ ,
\end{aligned}
\end{equation}

where $\s_0 \in \{g_0, \la_0, \x_0\}$ is any bare coupling, and $\s \in \{g, \la, \x\}$ is any renormalised coupling. Taking the logarithm of the coupling constants in \eqref{Bare couplings}, and only keeping terms that are quadratic in couplings yields
\begin{equation}
\begin{aligned}
\log \la_0 &= \e\log\m + \log\la + \frac{N_\ph + 8}{4\pi\e}\la + \frac{N_\ch}{4\pi\e}\frac{g^2}{\la} + ... \ , \\
\log \x_0 &= \e\log\m + \log\x + \frac{N_\ch + 8}{4\pi\e}\x + \frac{N_\ph}{4\pi\e}\frac{g^2}{\x} + ... \ , \\
\log g_0 &= \e\log\m + \log g + \frac{g}{\pi\e} + \frac{N_\ph + 2}{4\pi\e}\la + \frac{N_\ch + 2}{4\pi\e}\x + ... \ .
\end{aligned}
\end{equation}

Now differentiate these equations w.r.t. $\log\m$ and use the definitions \eqref{s-Definition Beta}
\begin{equation}
\begin{aligned}
\e + \frac{\be_\la}{\la} + \frac{N_\ph + 8}{4\pi\e}\be_\la + \frac{N_\ch}{4\pi\e}\frac{g}{\la} \left( 2\be_g - \frac{g}{\la}\be_\la \right) &= 0 \ , \\
\e + \frac{\be_\x}{\x} + \frac{N_\ch + 8}{4\pi\e}\be_\x + \frac{N_\ph}{4\pi\e}\frac{g}{\x} \left( 2\be_g - \frac{g}{\x}\be_\x \right) &= 0 \ , \\
\e + \frac{\be_g}{g} + \frac{\be_g}{\pi\e} + \frac{N_\ph + 2}{4\pi\e}\be_\la + \frac{N_\ch + 2}{4\pi\e}\be_\x &= 0 \ .
\end{aligned}
\end{equation}

The solution to these equations yields the $\be$-function
\begin{equation}
\begin{aligned}
\be_\la &= - \e\la + \frac{N_\ph + 8}{4\pi}\la^2 + \frac{N_\ch}{4\pi}g^2 + ... \ ,\\
\be_\x &= - \e\x + \frac{N_\ch + 8}{4\pi}\x^2 + \frac{N_\ph}{4\pi}g^2 + ... \ ,\\
\be_g &= - \e g + \frac{g^2}{\pi} + \frac{N_\ph + 2}{4\pi}\la g + \frac{N_\ch + 2}{4\pi}\x g + ... \ .
\end{aligned}
\end{equation}

\section{Functional determinants} \label{App: Scalar det}

In this appendix we path integrate a bosonic $O(N_\ph)\oplus O(N_\ch)$-vector that is massless in the bulk, but gains a tensor mass $m^{IJ}$ as it approaches the boundary. We will not assume any specific form of the boundary mass until it is needed. We will write the fluctuation correction to the boundary potential as 
\begin{equation} \label{Diff Op}
\begin{aligned}
V \supset \Ph^Im^{IJ}\Ph^J \ , \\
\end{aligned}
\end{equation}
\begin{equation}
\begin{aligned}
m^{IJ} &= \begin{pmatrix}
m_\ph^{ij} & g\ph_{cl}^j\ch_{cl}^b \\
g\ch_{cl}^a\ph_{cl}^k & m_\ch^{ab}
\end{pmatrix}^{IJ} \ , \\
m_\ph^{ij} &\equiv A^\la_g\de^{ij} + \la\hp_{cl}^i\hp_{cl}^j \ , \\ 
m_\ch^{ab} &\equiv A^g_\x\de^{ab} + \x\hc_{cl}^a\hc_{cl}^b \ .
\end{aligned}
\end{equation}

The constant $A^x_y$ can be found in \eqref{dual scalar masses}. In this appendix we will not use the exact form of $A^x_y$, although it is important that it is proportional to the coupling constants. Using the results of section \ref{Sec:Tr} (e.g. \eqref{eq:1lEff2}) we have
\begin{equation} \label{s-Trace2}
\begin{aligned}
 V_{\text{eff}}^{\text{1-loop}} &= \int_{\mathbb{R}^d_+}d^dxI_\mathcal{M} + \int_{\mathbb{R}^{d - 1}}d^{d - 1}x_\para I_{\p\mathcal{M}} \ , \\ 
I_\mathcal{M} &= \int_{\mathbb{R}^d}\frac{d^dk}{(2\pi)^d}\tr_{O(N)}\frac{\log[G^{IJ}(k)]}{2} \ , \\
I_{\p\mathcal{M}} &= \int_{\mathbb{R}^{d - 1}}\frac{d^{d - 1}k_\para}{(2\pi)^{d - 1}}\tr_{O(N)}\frac{\log[G^{IJ}_{b.c.}(k_\para)]}{2} \ .
\end{aligned}
\end{equation}

Here we trace over the $O(N)$-indices, $G^{IJ}$ is the momentum propagator in the bulk, and $G_{b.c.}^{IJ}$ is the momentum propagator in the boundary limit
\begin{equation} \label{Momentum Prop 3}
\begin{aligned}
G^{IJ}(k) &= \frac{\de^{IJ}}{k^2} \ , \quad G_{b.c.}^{IJ}(k_\para) = \left( m^{IJ} + |k_\para|\de^{IJ} \right)^{-1} \ .
\end{aligned}
\end{equation}

The logarithm of the bulk propagator is 
\begin{equation}
\begin{aligned}
\log[G^{IJ}(k)] &= -2\de^{IJ}\log|k| \ .
\end{aligned}
\end{equation}

This allows us to find $I_{\mathcal{M}}$ in \eqref{s-Trace2}. We will use spherical coordinates and regulate the divergences using a momentum cutoff $\La \gg 1$. The integral is on the form
\begin{equation} \label{s-Div Int}
\begin{aligned}
J_n(\La) \equiv \int_{0}^{\La}drr^{n - 1}\log(r) = \frac{\La^{n}}{n} \left( \log(\La) - \frac{1}{n} \right) \ .
\end{aligned}	
\end{equation}

Which yields
\begin{equation} \label{s-Bulk Int}
\begin{aligned}
I_{\mathcal{M}} &= -\frac{(N_\ph + N_\ch)S_d}{(2\pi)^d}\lim\limits_{\La\rightarrow\infty}J_d(\La) \ .
\end{aligned}	
\end{equation}

To compute $I_{\p\mcm}$ we will use that the logarithm of the inverse of a matrix can be expressed in terms the original matrix via
\begin{equation} \label{Log inverse}
\begin{aligned}
\left\{ \begin{array}{l l}
A &= e^{\log(A)} \quad\Rightarrow\quad A^{-1} = e^{-\log(A)} \\
A^{-1} &= e^{\log(A^{-1})}
\end{array} \right\}\Rightarrow\quad \log(A^{-1}) = - \log(A) \ .
\end{aligned}
\end{equation}

Using this we find the trace of the logarithm of the momentum propagator \eqref{Momentum Prop 3}
\begin{equation} \label{Log}
\begin{aligned}
&\log[G_{b.c.}^{IJ}(k_\para)] = -\log\left[|k_\para|\de^{IJ}\left(\frac{m^{IJ}}{|k_\para|} + \mbi \right)\right]  = -\de^{IJ}\log(|k_\para|) - \log\left(\frac{m^{IJ}}{|k_\para|} + \mbi \right) \ .
\end{aligned}
\end{equation} 

To find the second logarithm we diagonalise $m^{IJ}$. It has four eigenvalues. Two of these are
\begin{equation}
\begin{aligned}
\la_1 = A^\la_g \ , \quad \text{(with multiplicity $N_\ph - 1$),} \\
\la_2 = A^g_\x \ , \quad \text{(with multiplicity $N_\ch - 1$),}
\end{aligned}
\end{equation}

and the other two have both multiplicity one
\begin{equation}
\begin{aligned}
\la_\pm &= \frac{A^\la_g + \la\ph^2_{cl} + A^g_\x + \x\ch^2_{cl} \pm \sqrt{(A^g_\x + \x\ch^2_{cl} - A^\la_g - \la\ph^2_{cl})^2 + 4g^2\ph_{cl}^2\ch_{cl}^2}}{2} \ .
\end{aligned}
\end{equation}

We proceed with diagonalising the boundary mass $m^{IJ}$ using some matrix $A$ (as we will see, the exact form of $A$ does not matter)
\begin{equation} \label{Diag}
\begin{aligned}
m^{IJ} = (A^{-1}DA)^{IJ} \ , \quad D = \diag(\la_3^+, A^\la_g, ..., A^\la_g, \la_3^-, A^g_\x, ..., A^g_\x) \ .
\end{aligned}
\end{equation}

The second logarithm in \eqref{Log} can now be found from its Taylor expansion
\begin{equation} 
\begin{aligned}
&\log\left(\frac{m^{IJ}}{|k_\para|} + \mbi \right) = \sum_{n \geq 1}\frac{(-1)^{n + 1}}{n|k_\para|^n}((A^{-1}DA)^n)^{IJ} = (A^{-1})^{IK}\sum_{n \geq 1}\frac{(-1)^{n + 1}}{n|k_\para|^n}((D)^n)^{KL}A^{LJ} \\
\eq = \left( A^{-1}\diag\left(\log\left(\frac{\la_3^+}{|k_\para|} + 1\right), ...\right)A \right)^{IJ} \ . 
\end{aligned}
\end{equation}

Using cyclicity of the trace, we find
\begin{equation*} \hspace{-30px}
\begin{aligned}
&\tr\log[G_{b.c.}^{IJ}(k_\para)] = -\log\left(\la_+ + |k_\para|\right) - \log\left(\la_- + |k_\para|\right) - (N_\ph - 1)\log(A^\la_g + |k_\para|) - (N_\ch - 1)\log(A^g_\x + |k_\para|) \ .
\end{aligned}
\end{equation*}

The boundary integrals in \eqref{s-Trace2} are then on the form
\begin{equation}
\begin{aligned}
I_{\p\mathcal{M}} &= -\lim\limits_{\La\rightarrow\infty} \left[ K_{\la_3^+}(\La) + K_{\la_3^-}(\La) + (N_\ph - 1)K_{A^\la_g}(\La) + (N_\ch - 1)K_{A^g_\x}(\La) \right] \ , \\
K_x(\La) &= \frac{S_{d - 1}}{2^d\pi^{d - 1}}\int_{0}^{\La}drr^{d - 2} \log(x + r) \ .
\end{aligned}	
\end{equation}

This integral is a ${}_2F_1$-hypergeometric function. Its expansion in $\e$ in $3 - \e$ dimensions is performed using the HypExp mathematica package \cite{Huber:2005yg, Huber:2007dx}. We will keep terms up to order $\e^2$, assuming that the boundary mass is proportional to $\e$ ($A^x_y\propto\e$).\footnote{This assumption is motivated by the coupling constants at FP's in appendix \ref{App: Scalar beta}.} After this we expand around large $\La$, and neglect terms that goes as $\La^{-1}$
\begin{equation*}
\begin{aligned}
I_{\p\mathcal{M}} &= - \lim\limits_{\La\rightarrow\infty} \left[ \frac{N_\ph + N_\ch}{2^d\pi^{d - 1}}S_{d - 1}J_{d - 1}(\La) + \tilde{K}_\ph(\La) + \tilde{K}_\ch(\La) + \mco(\La^{-1}) \right] + \mco(\e^3) \ , \\
K_x(\La) &= -\frac{3d^2 - 22d + 43 - 2(d^2 - 8d + 19)}{16}\La^{d - 1}\log(\La) - (d - 4)\La^{d - 2}x + \frac{x^2}{2} \left( \log\left(\frac{x}{\La}\right) - \frac{1}{2} \right) \ .
\end{aligned}	
\end{equation*}

This, together with \eqref{s-Bulk Int} and \eqref{s-Trace2}, yields the full path integral over $\Phi$
\begin{equation}
\begin{aligned}
Z &= (2\pi)^{d/2} \exp\left[ -\lim\limits_{\La\rightarrow\infty} \left( \frac{N_\ph + N_\ch}{(2\pi)^d}S_dJ_d + \frac{N_\ph + N_\ch}{2^d\pi^{d - 1}}S_{d - 1}J_{d - 1} + \X_1 + \X_2 \right) \right] \ , \\
\end{aligned}
\end{equation}
\begin{equation}
\begin{aligned}
\X_1 &= -(d - 4)\La^{d - 2}(N_\ph A^\la_g + N_\ch A^g_\x + \la\ph_{cl}^2 + \x\ch_{cl}^2) \\
&= -(d - 4)\La^{d - 2}\frac{(N_\ch g + (N_\ph + 2)\la)\ph_{cl}^2 + (N_\ph g + (N_\ch + 2)\x)\ch_{cl}^2}{2} \ , \\
\X_2 &= \frac{N_\ph - 1}{2}(A^\la_g)^2 \left( \log\left(\frac{A^\la_g}{\La}\right) - \frac{1}{2} \right) + \frac{N_\ch - 1}{2}(A^g_\x)^2 \left( \log\left(\frac{A^g_\x}{\La}\right) - \frac{1}{2} \right) \\
\eq + \frac{\la_+^2}{2} \left( \log\left(\frac{\la_+}{\La}\right) - \frac{1}{2} \right) + \frac{\la_-^2}{2} \left( \log\left(\frac{\la_-}{\La}\right) - \frac{1}{2} \right) \ .
\end{aligned}
\end{equation}
By taking $N_\ph= N_\chi$ and $\xi=\lambda$ we obtain the result relevant for section \ref{Sec: Dual scalar CW}.

\bibliographystyle{utphys}
\footnotesize
\bibliography{paper}

\end{document}